\documentclass[12pt]{elsarticle}
\usepackage{graphicx}
\usepackage{graphicx,epstopdf}
\usepackage{caption}
\usepackage{subcaption}
\usepackage{float}
\usepackage{amssymb,amsmath}
\usepackage{exscale}
\usepackage{makeidx,shortvrb,latexsym}
\usepackage{epstopdf}
\usepackage{tabularx, booktabs, multirow}
\usepackage{array}
\usepackage[]{hyperref}
\usepackage{color}

\usepackage[bbgreekl]{mathbbol}

\usepackage{scalerel,stackengine}
\stackMath
\newcommand\reallywidehat[1]{%
\savestack{\tmpbox}{\stretchto{%
 \scaleto{%
   \scalerel*[\widthof{\ensuremath{#1}}]{\kern-.6pt\bigwedge\kern-.6pt}%
   {\rule[-\textheight/2]{1ex}{\textheight}}%WIDTH-LIMITED BIG WEDGE
 }{\textheight}%
}{0.5ex}}%
\stackon[1pt]{#1}{\tmpbox}%
}
\journal{Int. J. Eng. Sci.}
\newcounter{defcounter}
\begin{document}

\begin{frontmatter}
\title{DBFFT: A displacement based FFT approach for non-linear homogenization of the mechanical behavior}
\author{S. Lucarini$^{1, 2}$}
\author{J. Segurado$^{1, 2, }$\corref{cor1}}
\address{$^1$ IMDEA Materials Institute \\ C/ Eric Kandel 2, 28906, Getafe, Madrid, Spain. \\\ \\
$^2$ Department of Materials Science, Polytechnic University of Madrid/Universidad Polit\'ecnica de Madrid \\ E. T. S. de Ingenieros de Caminos. 28040 - Madrid, Spain.}
\cortext[cor1]{Corresponding author at: IMDEA Materials Institute, Spain \\ E-mail address: javier.segurado@imdea.org (J. Segurado)}

\begin{abstract}
 Most of the FFT methods available for homogenization of the mechanical response use the strain/deformation gradient as unknown, imposing their compatibility using Green's functions or projection operators. This implies the allocation of redundant information and, when the method is based in solving a linear equation, the rank-deficiency of the resulting system. In this work we propose a fast, robust and memory-efficient FFT homogenization framework in which the displacement field on the Fourier space is the unknown: the displacement based FFT (DBFFT) algorithm. The framework allows any general non-linear constitutive behavior for the phases and direct strain, stress and mixed control of the macroscopic load. In the linear case, the method results in a linear system defined in terms of linear operators in the Fourier space and that does not require a reference medium. The system has an associated full rank Hermitian matrix and can be solved using iterative Krylov solvers and allows the use of preconditioners. A preconditioner is proposed to improve the efficiency of the system resolution. Finally, some numerical examples including elastic, hyperelastic and viscoplastic materials are solved to check the accuracy and efficiency of the method. The computational cost reduction respect the Galerkin-FFT was around 30\%.
\end{abstract}

\begin{keyword}
FFT,  computational homogenization, micromechanics, polycrystals
\end{keyword}
\end{frontmatter}

\section{Introduction}

In computational homogenization and multiscale approaches, the efficiency solving the mechanical boundary value problem is crucial \citep{GEERS2010,SEGURADO2018}. More efficient methods allow the use of larger and more realistic Representative Volume Elements (RVEs) of the microstructure or to perform statistical analysis, fundamental for the study of fracture or fatigue. FFT based homogenization does not requiere meshing simplifying the pre-processing and
shows a remarkable computational efficiency and low memory allocation compared to Finite Element method (FE). The use of iterative methods is partially behind this improvement since FFT has been usually compared with classical implementations of FE that rely on less efficient direct solvers for the resulting linear systems. Nevertheless, even introducing iterative solvers and using regular grids in FE, FFT methods still much more efficient \cite{Lucarini2019a}. The memory reduction is due first to the fact that in FFT approaches, the linear operators do not need to be stored as a sparse matrix as normally done in FE. Second, the use of periodic boundary conditions in the FE by nodal elimination reduces drastically the sparsity of the stiffness matrix, resulting in additional memory storage and reducing the computational efficiency due to the increase of the number of operations when applying the linear operators. For these reasons, FFT homogenization has evolve from being a promising approach when it was introduced more than 20 years ago \cite{MOULINEC1994} to be a relatively standard technique used by tens of research groups working in homogenization.

There are many different approaches in FFT homogenization and, disregarding other possible classification, the methods can be divided in two groups attending to their formulations. The first group are methods derived from the original fixed point iteration scheme proposed by Moulinec and Suquet \citep{MOULINEC1994,MOULINEC1998}. These methods are derived from the strong formulation of the equilibrium and are based in the use of Green's functions for a reference medium computed in the Fourier space and the solution of the Lippman-Swinger equation. Several improvements of the original method were developed to improve the convergence of the iterative scheme for high stiffness contrast \citep{EYRE1999,MICHEL2000,Monchiet2012}. Alternatively, some researchers transformed the fixed-point iterative algorithms into a system of equations, solving the resulting system using Krylov solvers \citep{ZEMAN2010,BRISARD2010,BRISARD2012,KABLE2014,KABEL2016}. The second family is derived from the weak formulation of the linear momentum conservation \citep{VONDREJC2014,ZEMAN2017,DEGEUS2017,Lucarini2019a}. In this case, the Galerkin approach of the equilibrium is written in terms of the trigonometric polynomials, defined through the discrete Fourier Transform. In addition to its smart formulation, this approach can be highlighted because it does not need the definition of a reference medium. In this family, Krylov solvers are proposed as the kernel of the algorithm.

All these approaches present similarities and with particular choices of the reference mediums, methods which have a very different theoretical starting point, result in very similar algorithms (as an example the conjugate gradient method developed in \cite{KABLE2014} and Galerkin based methods). The most important common point of all the previous algorithms is the use of the strain field as unknown which compatibility is imposed using some Green's function or a projection operator.  If the algorithms are written as a linear system of equations, the strain field solution of the system is not unique (the system is rank-deficient) and the correct solution is obtained thanks to the ability of Krylov solvers to find the minimum norm solution.

Very recently, some authors \cite{Willot2014,SOK2016} have introduced an alternative view point, the use of the potential variable instead of its associated flux for the FFT homogenization of Poisson type problems. In \cite{Willot2014} the electrostatic potential was used as unknown instead of the electrical field while in \cite{SOK2016} this idea was extended to the homogenization of linear elastic heterogeneous media. However, in these seminal studies, attention is payed only to linear homogenization and to cases where the macroscopic value of the gradient of the potential field (electric field or displacement in the previous cases) is the variable imposed. Moreover, the Moulineq-Suquet fixed point iteration scheme was used as the algorithm behind the displacement based approaches, whose convergence properties are reduced compared to Krylov solvers methods \cite{ZEMAN2010}.

In this paper, a new full FFT homogenization framework for linear/non-linear mechanical problems is proposed based on the displacement field as the problem unknown and using the preconditioned conjugate gradient method as kernel, the Displacement Based FFT (DBFFT). In our approach, a system of linear equations in the Fourier space
is derived for the linear behavior with an associated full-rank Hermitian matrix, which is solved using a Krylov solver with a preconditioner. Moreover, the method allows imposing macroscopic values of strain, stress and mixed conditions maintaining the definiteness of the system and the computational performance. The non-linear extension of the method is totally general and can be used for any local constitutive equation and combination of macroscopic stress and strain history.

The paper is organized as follows. In section 2 the method will be derived for small strains and finite strains, and considering strain, stress and mixed control. In section 3 some numerical experiments will be proposed for different types of materials, covering from a matrix reinforced with elastic inclusions or voids to a visco-plastic polycrystals. In section 4, the numerical performance of the method will be analyzed and, finally, section 5 will present some conclusions.

\section{The method: Displacement Based FFT}
The method aims to find directly the displacement field that satisfies the equilibrium conditions. The properties of the Fourier Transform and the decomposition of the displacement field in its macroscopic part and the fluctuations will be used to solve the periodic boundary value problem. The problem consists in obtaining the displacement field that fulfills the equilibrium in a periodic domain $\Omega\in\mathbb{R}^3$, the Representative Volume Element (RVE). The RVE is a hexahedral box of dimensions $L_1 \cdot L_2 \cdot L_3$  formed by the spatial distribution of two or more phases, being the behavior at a point $ x\in\Omega$ defined by the constitutive equation of the phase occupying that point. The boundary conditions are periodic in strain and stress, and the load is introduced through a far-field/macroscopic state.

\subsection{Discretization}
The domain $\Omega$ is discretized in a voxelized regular grid containing $N_1\cdot N_2\cdot N_3$ voxels. The fields involved in the problem will then be represented by their value at the center of each voxel. The Fourier space is discretized in the same number of frequencies and the Fourier transform of a function defined in  $\Omega$ is obtained by the Discrete Fourier Transform of the discrete field and computed using the FFT algorithm. If $N_1,N_2$ and $N_3$ are taken as odd numbers, the Fourier space discretization is defined by the frequency vectors
\begin{equation}
\boldsymbol{\xi}=\xi_i=2\pi \frac{n_i-(N_i+1)/2}{L_i}\quad \text{ for } \quad n_i =1, \dots, N_i \quad ,
\end{equation}
where $n_i$ is the frequency number and $L_i$ and $N_i$ are, respectively, the length of the domain $\Omega$ in the $i-th$ direction and the number of points in which is discretized. The use of even grid implies some assumptions related to the Nyquist frequencies such as neglect them or approximate them. For the shake of simplicity, in this work only odd grids will be analyzed.
\subsection{DBFFT scheme: Small strains}
Under macroscopic strain control loading, the prescribed macroscopic strain tensor $\overline{\boldsymbol{\varepsilon}}_{\overline{U}}$ can be written as function of the relative displacement vectors between periodic faces, $\overline{\mathbf{U}}$, as
\begin{equation}\label{eq:disp_strain}
\left[\overline{\varepsilon}_{\overline{U}}\right]_{ij}=\frac{1}{2}\left(\frac{\overline{U}_{i(j)}}{L_j}+\frac{\overline{U}_{j(i)}}{L_i}\right)
\end{equation}

where $\overline{U}_{i(j)}$ is the relative displacement between the periodic faces normal to direction $j$, in the direction $i$ being $L_j$ the length of the domain $\Omega$ along the direction $j$. Note that this relation is the same used for periodic boundary conditions in Finite Element homogenization, in which displacements are also the principal unknown \cite{SEGURADO2018}

The displacement field is split into its fluctuating periodic part $\widetilde{\mathbf{u}}\left(\mathbf{x}\right)$ (with zero average), and a linearly growing part, $\mathbf{u}_{\overline{U}}$, that accounts for the macroscopic strain and is dictated by the relative displacements vectors $\overline{\mathbf{U}}$,
\begin{equation}\label{eq:disp_decomp}
\mathbf{u}\left(\mathbf{x}\right) =\mathbf{u}_{\overline{U}}\left(\mathbf{x}\right)+\widetilde{\mathbf{u}}\left(\mathbf{x}\right).
\end{equation}
The displacement field contribution due to the macroscopic relative displacement is related to the prescribed macroscopic strain tensor by
\begin{equation}\label{eq:decomp_displ}
\mathbf{u}_{\overline{U}}\left(\mathbf{x}\right)=\overline{\boldsymbol{\varepsilon}}_{\overline{U}}\cdot\mathbf{x}
\end{equation}
Using this decomposition (eq. \ref{eq:disp_decomp}), the resulting expression for the strain tensor field $\boldsymbol{\varepsilon}\left(\mathbf{x}\right)$ reads as
\begin{equation}\label{eq:strain_decomp}
\boldsymbol{\varepsilon}\left(\mathbf{x}\right)=\nabla^s\mathbf{u}\left(\mathbf{x}\right) =\overline{\boldsymbol{\varepsilon}}_{\overline{U}}+\nabla^s\widetilde{\mathbf{u}}\left(\mathbf{x}\right)
\end{equation}
where $\nabla^s$denotes the symmetric gradient.

The starting point of the method is the equilibrium equation in its strong form,
\begin{equation}\label{eq:strong1}
\nabla\cdot\boldsymbol{\sigma}\left(\mathbf{x}\right)=\mathbf{0}.
\end{equation}
If the phases inside the domain are linear elastic materials, the equation (\ref{eq:strong1}) can be written as
\begin{equation}\label{eq:eqlin}
\nabla\cdot \left[ \mathbb{C}\left(\mathbf{x}\right):\nabla^s \mathbf{u}\left(\mathbf{x}\right)\right]=\mathbf{0}
\end{equation}
where $\mathbb{C}\left(\mathbf{x}\right)$ corresponds to the stiffness tensor of the phase located at the point $\mathbf{x}$.
Using the displacement field decomposition (eq. \ref{eq:disp_decomp}), the linearity of both the gradient and divergence differential operators and rearranging terms, the equilibrium equation can be rearranged as
\begin{equation}\label{eq:system1}
\nabla\cdot \left[ \mathbb{C}\left(\mathbf{x}\right):\nabla^s \widetilde{\mathbf{u}}\left(\mathbf{x}\right)\right]=
-\nabla\cdot \left[ \mathbb{C}\left(\mathbf{x}\right):\overline{\boldsymbol{\varepsilon}}_{\overline{U}}\right]
\end{equation}
Equation (\ref{eq:system1}) is a partial differential equation in which the displacement fluctuation, $\widetilde{\mathbf{u}}$, is the field to solve. This equation is also the starting point of the Moulineq and Suquet scheme \cite{MOULINEC1994} but, instead of introducing a reference medium to transform the microstructure dependency $\mathbb{C}(\mathbf{x})$ into an eigenstrain, the equation is transformed in its current form to the Fourier space to compute the derivatives. Note that the Fourier transform of the derivative of a function $f(x):\mathbb{R}\rightarrow\mathbb{R}$ is defined as
\begin{equation}
\mathcal{F}\left(\frac{\mathrm{d}}{\mathrm{d}x}f(x)\right)=\mathrm{i} \xi \mathcal{F}\left(f(x)\right)
\label{eq:1Ddif}
\end{equation}
where $\mathcal{F}$ is the  Fourier transform of a real valued function, $\xi$ is the Fourier spatial frequency and $\mathrm{i}$  represents the imaginary unit.

Using the definition of the derivative in the Fourier space (eq. \ref{eq:1Ddif}), the symmetric gradient of the fluctuation field in the real space
\begin{equation}\label{eq:symgrad0}
\left[\nabla^s \widetilde{\mathbf{u}}\left(\mathbf{x}\right)\right]_{ij}=\frac{1}{2}\left(\frac{\partial \widetilde{u}_i}{\partial x_j}+\frac{\partial \widetilde{u}_j}{\partial x_i}\right)
\end{equation}
is transformed to
\begin{equation}\label{eq:symgrad}
\left[\mathcal{F}\left(\nabla^s \widetilde{\mathbf{u}}\left(\mathbf{x}\right)\right)\right]_{ij}=\widehat{s}_{ijk}\widehat{\widetilde{u}}_k=\widehat{\mathbb{s}}\left(\boldsymbol{\xi}\right)\cdot\widehat{\widetilde{\mathbf{u}}}\left(\boldsymbol{\xi}\right)
\end{equation}
being $\widehat{\mathbb{s}}$ the symmetric gradient operator and $\widehat{\widetilde{\mathbf{u}}}$ the displacement field, both fields belonging to the Fourier space and therefore defined in the frequency domain as function of the frequency vector $\boldsymbol{\xi}$. The expression of the symmetric gradient operator $\widehat{\mathbb{s}}$ is
\begin{equation}
\widehat{\mathbb{s}}(\boldsymbol{\xi})=\widehat{s}_{ijk}(\boldsymbol{\xi})=\frac{1}{2}\left( i\xi_j \delta_{ik}+i\xi_i \delta_{jk}\right)
\end{equation}
and it can be observed that the operator becomes zero for the null frequency $\boldsymbol{\xi}=\mathbf{0}$.

The divergence of a tensor field is defined in the real space as
\begin{equation}\label{eq:diver0}
\left[\nabla\cdot \boldsymbol{\sigma}\left(\mathbf{x}\right)\right]_{i}=\frac{\partial \sigma_{ij}}{\partial x_j}
\end{equation}
and its transformation in the Fourier space corresponds to
\begin{equation}\label{eq:diver}
\left[\mathcal{F}\left(\nabla \cdot \boldsymbol{\sigma}\left(\mathbf{x}\right)\right)\right]_{i}=\widehat{d}_{ijk}\widehat{\sigma}_{jk}=\widehat{\mathbb{d}}\left(\boldsymbol{\xi}\right):\widehat{\boldsymbol{\sigma}}\left(\boldsymbol{\xi}\right)
\end{equation}
being $\widehat{\mathbb{d}}$ the divergence operator and $\widehat{\boldsymbol{\sigma}}$ the stress tensor field, both expressed in the frequency domain. The expression of the operator $\widehat{\mathbb{d}}$ corresponds to
\begin{equation}
\widehat{\mathbb{d}}=\widehat{d}_{ijk}=i\xi_k\delta_{ij}
\end{equation}
and it also eliminates the average part of the field.

Finally, transforming the equation linear momentum conservation (eq.  \ref{eq:system1}) to the Fourier space and replacing the differential operators in the real space by their Fourier space counterparts, given by eqs. (\ref{eq:symgrad}) and (\ref{eq:diver}), it is obtained
\begin{equation}\label{eq:lineqf}
\widehat{\mathbb{d}}: \mathcal{F}\left( \mathbb{C}(\mathbf{x}):\mathcal{F}^{-1}\left(\widehat{\mathbb{s}}\cdot\widehat{\widetilde{\mathbf{u}}}\right)\right)=-\widehat{\mathbb{d}}:\mathcal{F}\left( \mathbb{C}(\mathbf{x}):\overline{\boldsymbol{\varepsilon}}_{\overline{U}}\right)
\end{equation}
If the fields in the previous equation are replaced by their discrete counterparts in both real and Fourier spaces and the Fourier transform is computed as the discrete transform, equation \ref{eq:lineqf} becomes a linear system of equations of complex numbers in which the unknown is the fluctuation displacement field defined in the Fourier space, $\widetilde{\mathbf{u}}$. The size of the system in 3D is $3\cdot N_1\cdot N_2 \cdot N_3$ and, if the zero frequency terms and real Fourier transform symmetries are removed, the system becomes fully determined  with an associated Hermintian matrix. These type of linear systems can be solved with direct or iterative methods. In the  case of iterative solvers, the full-rank of the associated matrix allows the use of preconditioners, in contrast to what happens when solving the rank deficient systems resulting from the Galerkin-FFT approach \cite{VONDREJC2014,ZEMAN2017}

\subsection{DBFFT: Stress control}
Many times the macroscopic loading path is prescribed in stress or with a combination of terms in stress and strains. Classically, FFT methods are conceived to have the macroscopic strain as input and resolve the stress and mixed  cases by iterative approaches. Very recently, the authors have proposed a method to include directly the prescribed terms of the macroscopic stress in the formulation \cite{Lucarini2019b} for the Galerkin-FFT approach. In this paper, we propose a similar approach for the DBFFT.

The macroscopic load should be imposed by six different values of either the macroscopic strain or stress tensor. Let $\left[\overline{\sigma}_{\overline{f}}\right]_{IJ}$ be the terms of the macroscopic stress tensor that are prescribed and let $[\overline{\boldsymbol{\varepsilon}}_{\overline{U}}]_{ij}$  be the terms prescribed of the macroscopic strain, with  with $IJ\neq ij$. Equivalent to the treatment of the macroscopic strain, the macroscopic stress is imposed through a force vector $\bar{\mathbf{f}}$
\begin{equation}\label{eq:disp_stress}
\left[\overline{\sigma}_{\overline{f}}\right]_{ij}=\frac{\overline{f}_{i(j)}}{A_j}
\end{equation}
where $\overline{f}_{i(j)}$ is the force in $i$ direction applied on the face normal to $j$ direction and $A_j$ the area of the $\Omega$ domain a the plane normal to $j$ direction. Equilibrium implies that the same forces appear with contrary sign in the opposite directions. The microscopic stress in the terms where the macroscopic stress is prescribed are split in its average part and fluctuations, while the rest of the terms of the tensor are not decomposed . This decomposition is given by
\begin{eqnarray}\label{eq:stress_decomp2}
[\boldsymbol{\sigma}\left(\mathbf{x}\right)]_{IJ}=[ \overline{\boldsymbol{\sigma}}_{\overline{f}} ]_{IJ}+
[ \boldsymbol{\sigma}^{\ast}\left(\mathbf{x}\right)]_{IJ} \\ \nonumber
\left[\boldsymbol{\sigma}\left(\mathbf{x}\right)\right]_{ij}=[\boldsymbol{\sigma}^{\ast}\left(\mathbf{x}\right)]_{ij}
\end{eqnarray}

In equation (\ref{eq:stress_decomp2}) $\boldsymbol{\sigma}^{\ast}\left(\mathbf{x}\right)$ is a non-zero averaged tensor in the components where strain is imposed, $ij$, and a stress fluctuation (zero average) in the components $IJ$ where stress is imposed. The  strain field can be split as
\begin{equation}\label{eq:strain_decomp2}
\boldsymbol{\varepsilon}\left(\mathbf{x}\right)=\nabla^s\mathbf{u}\left(\mathbf{x}\right)=
%\overline{\boldsymbol{\varepsilon}}_{\overline{f}}%+\nabla^s\mathbf{u}^{\ast}\left(\mathbf{x}\right)=
\overline{\boldsymbol{\varepsilon}}_{\overline{U}}+\overline{\boldsymbol{\varepsilon}}_{\overline{f}}+\nabla^s\widetilde{\mathbf{u}}\left(\mathbf{x}\right)
\end{equation}
where $\overline{\boldsymbol{\varepsilon}}_{\overline{f}}$ is the unknown macroscopic strain that appears as a result of the macroscopic prescribed stress $\overline{\boldsymbol{\sigma}}_{\overline{f}}$, and that is zero in the $ij$ components. The tensor $\overline{\boldsymbol{\varepsilon}}_{\overline{U}}$ contains the prescribed components of the macroscopic strain $ij$ and is zero in the components $IJ$. The displacement field corresponding to the strains defined in eq. (\ref{eq:strain_decomp2}) is given by
\begin{equation}\label{eq:disp_decomp2}
\mathbf{u}\left(\mathbf{x}\right) =\overline{\boldsymbol{\varepsilon}}_{\overline{U}}\cdot\mathbf{x}+\overline{\boldsymbol{\varepsilon}}_{\overline{f}}\cdot\mathbf{x} + \widetilde{\mathbf{u}}\left(\mathbf{x}\right)
\end{equation}

The displacement field contains now two different unknonwns, $\overline{\boldsymbol{\varepsilon}}_{\overline{f}}$ and $\nabla^s\widetilde{\mathbf{u}}$. This implies the introduction of additional equations to impose that the $IJ$ components of the average stress equal the corresponding components of the prescribed macroscopic stress. The resulting system of differential equations reads
\begin{eqnarray}\label{eq:strong2stress}
\begin{aligned}
\nabla\cdot\boldsymbol{\sigma}\left(\mathbf{x}\right)=\mathbf{0}\\
\left[\left<\boldsymbol{\sigma}\left(\mathbf{x}\right)\right>\right]_{IJ}=\left[\overline{\boldsymbol{\sigma}}_{\overline{f}}\right]_{IJ}
\end{aligned}
\end{eqnarray}
where the unknown variable to solve is $\left\{\widetilde{\mathbf{u}}\ |\ \overline{\boldsymbol{\varepsilon}}_{\overline{f}}\right\}$. Note that the first equation does not affect the average value, and the second are the additional equations needed to obtain the unknown $IJ$ components of the macroscopic strain  $\overline{\boldsymbol{\varepsilon}}_{\overline{f}}$. Again for simplicity, the phases will be considered linear elastic materials, so that, using the displacement field decomposition (eq. \ref{eq:disp_decomp2}), equilibrium equation becomes
\begin{equation}\label{eq:strong4stress}
\nabla\cdot \left[ \mathbb{C}\left(\mathbf{x}\right):\nabla^s \left(\widetilde{\mathbf{u}}\left(\mathbf{x}\right)+\overline{\boldsymbol{\varepsilon}}_{\overline{F}}\cdot\mathbf{x}+\overline{\boldsymbol{\varepsilon}}_{\overline{U}}\cdot\mathbf{x}\right)\right]=0.
\end{equation}
Then, thanks to the linearity of gradient and divergence operators the terms are rearranged, yielding into expression (\ref{eq:strong5stress}).
\begin{equation}\label{eq:strong5stress}
\nabla\cdot \left[ \mathbb{C}\left(\mathbf{x}\right):\left(\nabla^s \widetilde{\mathbf{u}}\left(\mathbf{x}\right)+\overline{\boldsymbol{\varepsilon}}_{\overline{F}}\right)\right]=-\nabla\cdot \left[ \mathbb{C}\left(\mathbf{x}\right):\overline{\boldsymbol{\varepsilon}}_{\overline{U}}\right]
\end{equation}
The system is transformed to the Fourier space using the same differential operators defined in the previous section. The terms inside divergence operations in eq. (\ref{eq:strong5stress}), the left-hand side (lhs) and the right-hand side (rhs), are rewritten in the Fourier space upon the use of eq. \ref{eq:symgrad}.
\begin{eqnarray}
\begin{aligned}
\widehat{\boldsymbol{\sigma}}_{lhs}=\mathcal{F}\left(\mathbb{C}:\left(\mathcal{F}^{-1}\left(\widehat{\mathbb{s}}\cdot\widehat{\widetilde{\mathbf{u}}}\right)+\overline{\boldsymbol{\varepsilon}}_{\overline{F}}\right)\right)\\
\widehat{\boldsymbol{\sigma}}_{rhs}=\mathcal{F}\left(\mathbb{C}:\overline{\boldsymbol{\varepsilon}}_{\overline{U}}\right)
\end{aligned}
\end{eqnarray}
Finally, using eq. (\ref{eq:diver}), the  system of equations defining the equilibrium (\ref{eq:strong2stress}) becomes
\begin{eqnarray}
\begin{aligned}
\widehat{\mathbb{d}}:\widehat{\boldsymbol{\sigma}}_{lhs}=-\widehat{\mathbb{d}}: \overline{\boldsymbol{\sigma}}_{rhs}\\
\left[\widehat{\boldsymbol{\sigma}}_{lhs}\left(\mathbf{0}\right)\right]_{IJ}=\left[\overline{\boldsymbol{\sigma}}_{\overline{f}}\right]_{IJ}-\left[\widehat{\boldsymbol{\sigma}}_{rhs}\left(\mathbf{0}\right)\right]_{IJ}
\end{aligned}
\end{eqnarray}
As in the strain controlled version, the system of equations is fully determined and Hermitian if the zero frequency from the first equation is removed and the symmetries of the real Fourier transform are used.

\subsection{DBFFT: Non-linear problems}
The extension of the method to finite strains is straightforward expressing the equilibrium in the reference configuration through the first Piola Kirchhoff stress $\mathbf{P}$ and using the deformation gradient $\mathbf{F}$ to characterize the deformed state. For strain control, the prescribed macroscopic state is given by the average deformation gradient $\overline{\mathbf{F}}_{\overline{U}}$ that is related with the relative displacement between opposite RVE faces through
\begin{equation}\label{eq:disp_F}
\left[\overline{\mathbf{F}}_{\overline{U}}\right]_{ij}=\delta_{ij}+\frac{\overline{U}_{i(j)}}{L_j}.
\end{equation}
where $\delta_{ij}$ is the Dirac delta and  $\overline{U}_{i(j)}$ is the relative displacement between the periodic faces normal to direction $j$, in the direction $i$.

The conservation of the linear momentum in the reference configuration is given by
\begin{equation}\label{eq:euler-lagrangefinite0}
\nabla_0\cdot\mathbf{P}\left(\mathbf{x}\right)=\mathbf{0}
\end{equation}
where the first Piola-Kirchhoff stress $\mathbf{P}$ is given for each phase present in the microstructure by a non-linear constitutive equation  $\mathbf{P}(\mathbf{F},\mathbf{L},\boldsymbol{\alpha})$ depending in general on the deformation gradient, the velocity gradient $,\mathbf{L}$ and a set of internal variables $\boldsymbol{\alpha}$.  $\nabla_0\cdot$ represents the divergence in the reference configuration. As in the linear case, the variable in which the problem is defined is the displacement field, in particular the displacement fluctuations $\widetilde{\mathbf{u}}$.

The total macroscopic deformation gradient prescribed is split in $n$ increments. The deformation at each increment $k$, $\bar{\mathbf{F}}^{k}_{\overline{U}}$, defines the macroscopic relative displacement $\overline{\mathbf{U}}^{k} $ of the RVE opposite faces and therefore the contribution of this macroscopic displacement into the local field. If all the fields are known at increment $k$, the displacement at increment $k+1$ corresponds to
\begin{equation}\label{eq:decomp_displ2}
\mathbf{u}^{k+1}\left(\mathbf{x}\right) = (\bar{\mathbf{F}}_{\bar{U}}^{k+1}-\mathbf{I})\cdot\mathbf{x}+\widetilde{\mathbf{u}}^{k+1}\left(\mathbf{x}\right).
\end{equation}
where $\widetilde{\mathbf{u}}^{k+1}\left(\mathbf{x}\right)$ is the unknown field to be solved that corresponds to the fluctuation part of the displacement at increment $k+1$. The expression for the displacement (eq. \ref{eq:decomp_displ2}) is introduced in the equilibrium equation leading to
\begin{equation}\label{eq:euler-lagrangefinite}
\nabla_0\cdot\mathbf{P}(\mathbf{I}+\nabla_0 (\mathbf{u}^{k+1}\left(\mathbf{x}\right)))=\nabla_0\cdot\mathbf{P}(\bar{\mathbf{F}}_{\bar{U}}^{k+1}+\nabla_0 \widetilde{\mathbf{u}}^{k+1}\left(\mathbf{x}\right) )=\mathbf{0}
\end{equation}
where, the gradient of the deformation is $\nabla_0$ is also taken in the reference configuration, being $\mathbf{F}=\mathbf{I}+\nabla_0 \mathbf{u}$. Note that a non-linear elastic material has been assumed for simplicity (no dependency on $\mathbf{L}$ and $\boldsymbol{\alpha}$). The non-linear differential equation defined in eq. (\ref{eq:euler-lagrangefinite}) is solved iteratively by Newton method. For this purpose the displacement fluctuation field and the Piola stress are linearized around the last iteration $i$ of the displacement field fluctuations, $\widetilde{\mathbf{u}}^{k+1,i}\left(\mathbf{x}\right)$, named $\widetilde{\mathbf{u}}^{i}$ to alleviate the notation.
\begin{eqnarray}
\widetilde{\mathbf{u}}^{i+1}=\widetilde{\mathbf{u}}^{i}+\delta \widetilde{\mathbf{u}} \\
\mathbf{P}(\mathbf{I}+\nabla_0 (\mathbf{u}^{k+1}\left(\mathbf{x}\right))  )
\approx\mathbf{P}(\bar{\mathbf{F}}_{\bar{U}}^{k+1}+\nabla_0 \widetilde{\mathbf{u}}^{i})+\mathbb{K}^{i}:\nabla_0 \delta \widetilde{\mathbf{u}}\ \text{ with }\ \mathbb{K}^{i}=\frac{\partial \mathbf{P}}{\partial \nabla_0 \delta \widetilde{\mathbf{u}} }
\label{eq:lineariz}
\end{eqnarray}
Combining the equilibrium equation (\ref{eq:euler-lagrangefinite}) with the linearization of the stress (eq. \ref{eq:lineariz}), the next equation is derived
\begin{equation}\label{eq:eqfinite}
\nabla_0\cdot  \mathbb{K}^{i}: \nabla_0\delta\widetilde{\mathbf{u}}=-\nabla_0 \cdot\mathbf{P}\left(\mathbf{I}+\nabla_0 \mathbf{u}^{i}\right)
\end{equation}
that is a linear differential equation which unknown is $\delta\widetilde{\mathbf{u}}$. In the equation $\nabla_0 \mathbf{u}^{i}=\bar{\mathbf{F}}_{\bar{U}}^{k+1}-\mathbf{I}+\nabla_0 \widetilde{\mathbf{u}}^{i}$ and $\mathbb{K}^{i}$ is the material consistent tangent evaluated in the $i$-th iteration of increment $k+1$. The iterative Newton starts with $i=1$ and $\nabla_0\widetilde{\mathbf{u}}^{0}=\nabla_0\widetilde{\mathbf{u}}^{k}$.

Then, following the same approach as in the linear case, the linear differential equation is transformed to the Fourier space and the fields are replaced by their discrete counterparts. The result is a linear system of complex numbers
\begin{equation}\label{eq:nonlineqf}
\widehat{\mathbb{d}}: \mathcal{F}\left(\mathbb{K}^{i}:\left(\mathcal{F}^{-1}\left(\widehat{\mathbb{g}}\cdot\delta\widehat{\widetilde{\mathbf{u}}}\right)\right)\right)=-\widehat{\mathbb{d}}:\mathcal{F}\left( \mathbf{P}\left(\nabla_0\mathbf{u}^{i}+\mathbf{I}\right)\right)
\end{equation}
in which the unknown vector to solve is $\delta\widetilde{\mathbf{u}}$. The linear equation to solve the equilibrium in each Newton iteration (\ref{eq:nonlineqf}) is formally identical to the linear equation obtained for the case of linear elasticity (eq. \ref{eq:lineqf}). The difference is that in eq. (\ref{eq:nonlineqf}) the gradient operator $\widehat{\mathbb{g}}$ is no longer symmetric being this operator defined in the Fourier space as
\begin{equation}
\left[\widehat{\mathbb{g}}\left(\boldsymbol{\xi}\right)\right]_{ijk}=\delta_{ik}\xi_j.
\end{equation}
The resulting system is fully determined removing the zero frequency. As in the other cases, the linear equation can be solved by any iterative solver. The Newton iterations finish when the maximum value of $\nabla_0\delta\widetilde{\mathbf{u}}$ the last linear iteration  is below a tolerance set to $10^{-6}$ .

\subsubsection*{Stress and mixed control}
The stress and mixed control approach followed for the linear elastic problem can be followed almost identically in the non-linear case. The prescribed data in this case is a combination of some components the macroscopic first Piola-Kirchhoff stress $[\bar{\mathbf{P}}_{\overline{f}}]_{IJ}$ and other components of the macroscopic deformation gradient $[\bar{\mathbf{F}_{\overline{U}}}]_{ij}$. The deformation gradient field will be split into the part due to macroscopic imposed Piola stress $\overline{\mathbf{F}}_{\overline{f}}$ and the part due to prescribed deformation gradient $\overline{\mathbf{F}}_{\overline{U}}$.
\begin{equation}
\mathbf{F}\left(\mathbf{x}\right)=\mathbf{I}+\left[\overline{\mathbf{F}}_{\overline{U}}-\mathbf{I}\right]_{ij}+\left[\overline{\mathbf{F}}_{\overline{f}}-\mathbf{I}\right]_{IJ}+\nabla_0\widetilde{\mathbf{u}}
\end{equation}
where $\overline{\mathbf{F}}_{\overline{f}}$ is unknown and is only valued in $IJ$ components where Piola stress is imposed.
Dividing the applied load in $N$ increments and solving linearization of the equilibrium in each increment by an iterative Newton approach, the linear system to solve at each Newton iteration $i$ is
\begin{eqnarray}\label{eq:nonlineqfstress}
\widehat{\mathbb{d}}: \mathcal{F}\left(\mathbb{K}{i}:\left(\mathcal{F}^{-1}\left(\widehat{\mathbb{g}}\cdot\delta\widehat{\widetilde{\mathbf{u}}}\right)+\left(\overline{\mathbf{F}}_{\overline{f}}-\mathbf{I}\right)\right)\right)=\nonumber \\
=-\widehat{\mathbb{d}}:\mathcal{F}\left( \mathbf{P}\left(\nabla_0\mathbf{u}^{i}+\mathbf{I}\right)\right) \nonumber \\
\left[\mathcal{F}\left(\mathbb{K}{i}:\left(\mathcal{F}^{-1}\left(\widehat{\mathbb{g}}\cdot\delta\widehat{\widetilde{\mathbf{u}}}\right)+\left(\overline{\mathbf{F}}_{\overline{f}}-\mathbf{I}\right)\right)\right)\left(\mathbf{0}\right)\right]_{IJ}= \nonumber \\
=\left[\bar{\mathbf{P}}\right]_{IJ}-\mathcal{F}\left( \mathbf{P}\left(\nabla_0\mathbf{u}^{i}+\left[\mathbf{I}\right)\right)\left(\mathbf{0}\right)\right]_{IJ}
\end{eqnarray}
being $\nabla_0 \mathbf{u}^{i}=\left[\bar{\mathbf{F}}_{\bar{U}}^{k+1}-\mathbf{I}\right]_{ij}+\nabla_0 \widetilde{\mathbf{u}}^{i}$ and starting the algorithm with $\nabla_0\widetilde{\mathbf{u}}^{0}=\nabla_0\widetilde{\mathbf{u}}^{k}$. In \ref{eq:nonlineqfstress}, the unknown to be solved is the variable $\left\{\widehat{\widetilde{\mathbf{u}}}\ |\ \overline{\mathbf{F}}_{\overline{f}}\right\}$ and, if the zero frequency and doubled terms due to the symmetries of the real Fourier transform are removed in the first equation, the system becomes also fully determined and Hermitian.

\subsection{Improving the efficiency: Preconditioners}
The idea of introducing preconditioners to improve the convergence rate when using Krylov solvers for FFT homogenization was first proposed in \cite{ZEMAN2010} for the conjugate-gradient version of the basic-scheme applied to an electrostatic problem. However, its practical use has been limited since the application of a pre-conditioner to a rank-deficient system solved by the conjugate gradient, as the one resulting in the Galerkin-FFT approach, leads to either non convergence of the algorithm or to solutions different to the minimum-norm one.

On the contrary, in the DBFFT approach, thanks to full rank of the linear systems of equations derived for both linear (eq. \ref{eq:lineqf}) and non-linear materials (eq. \ref{eq:nonlineqf}), preconditioners can be applied without any risk in order to improve the convergence when solving iteratively the system.

The linear systems defined in the previous sections are not given as function of a matrix of coefficients but expressed in terms of linear operators as
\begin{equation}
\mathcal{A}(\mathbf{u})=\mathbf{b}
\end{equation}
where the linear operator $\mathcal{A}$ defines the result of applying the matrix of the system to a vector $\mathbf{u}$ and $\mathbf{b}$ is the right hand side of the equation.

In terms of linear operators, the preconditioner of the system should be a linear operator $\mathcal{M}$ that provides a good approximation of the inverse of the linear operator $\mathcal{A}$.
\begin{equation}\label{eq:pred1}
\mathcal{M}\left(\mathcal{A}\left(\mathbf{u}\right)\right)\approx\mathbf{u}\text{  and  }\mathcal{M}\approx\mathcal{A}^{-1}
\end{equation}
If an operator $\mathcal{M}$ were obtained which corresponded exactly to the inverse $\mathcal{A}^{-1}$ then the system could be directly solved applying $\mathbf{u}=\mathcal{M}(\mathbf{b})=\mathcal{A}^{-1}(\mathbf{b})$. For the systems considered it is not possible to obtain an analytic expression for $\mathcal{A}^{-1}$.

Focusing on the linear case given by eq. (\ref{eq:lineqf}), the inverse of $\mathcal{A}$ would consist in applying from left to right the inverse of the operations defining the action of $\mathcal{A}$ over $\widehat{\widetilde{\mathbf{u}}}$. Some of these operations are not invertible but, if the local stiffness matrix $\mathbb{C}(\mathbf{x})$ is replaced at each point by the average stiffness $\overline{\mathbb{C}}=1/n \sum_{\mathbf{x}_1}^{\mathbf{x}_n} \mathbb{C}\left(\mathbf{x}\right)$,  the modified eq. (\ref{eq:lineqf}) can be inverted. The resulting linear operator is given by
\begin{equation}\label{eq:pred2}
[\mathcal{M}(\widehat{\widetilde{\mathbf{u}}})]_{\boldsymbol{\xi}}:=\mathbf{M}(\boldsymbol{\xi})\cdot[\widehat{\widetilde{\mathbf{u}}}]_{\boldsymbol{\xi}}
\end{equation}
where $\widehat{\widetilde{\mathbf{u}}}$ is the discrete vector field of displacement fluctuations in the Fourier space and $[\widehat{\widetilde{\mathbf{u}}}]_{\boldsymbol{\xi}}$ is the displacement fluctuation vector in the Fourier point $\boldsymbol{\xi}$ and $\mathbf{M}(\boldsymbol{\xi})$ is second order tensor  defined for each frequency as
\begin{equation}\label{eq:pred3}
\mathbf{M}\left(\boldsymbol{\xi}\right)=\left\{\begin{array} {lr}\left[\boldsymbol{\xi}\cdot\overline{\mathbb{C}}\cdot\boldsymbol{\xi}\right]^{-1}&\text{for }\quad \boldsymbol{\xi}\ne \mathbf{0}\\\left[\left\{\mathbf{1}\right\}\cdot\overline{\mathbb{C}}\cdot\left\{\mathbf{1}\right\}\right]^{-1}&\text{for }\quad \boldsymbol{\xi}=\mathbf{0}\end{array}\right.
\end{equation}
The second order preconditioner operator is computed once at the beginning of the simulation and does not imply any extra direct or inverse Fourier transform. In the case of a finite strain non-linear simulation, the definition of the preconditioner $\mathbb{M}$ is very similar, replacing the average stiffness matrix by the average material tangent $\overline{\mathbb{K}}$ and it is computed at the beginning of the increment. This average stiffness matrix can be recomputed once in a while to further improve the convergence.

\section{Numerical results}

The DBFTT method has been added to the code FFTMAD \cite{Lucarini2019a, Lucarini2019b} as a new scheme and programmed combining python and fortran subroutines. In particular, the constitutive equations are programmed using abaqus UMATs standard and the evaluation of the constitutive equations at each iteration are performed using a fortran routine parallelized using {\em openMP}.All the linear systems are solved using the conjugate gradient method for complex numbers and the preconditioner defined in eqs. (\ref{eq:pred2}) and (\ref{eq:pred3}).  The conjugate gradient method for complex numbers has been programmed in python to compute the three residuals proposed at each equilibrium iteration. Postprocessing is done using FFTMAD subroutines and deformed plots and field iso plots are generated using {\em paraview}.

In this section several numerical examples are tested and compared with the FFT variational approach \citep{ZEMAN2017,DEGEUS2017} to benchmark the method proposed.
Three types of residual are computed to check the fulfilment of equilibrium condition, strain field compatibility and boundary conditions. The residuals proposed are similar to the ones proposed in \citep{MOULINEC2014} adapted to displacements.

The first residual defines the deviation of the resulting displacement field from the equilibrium, given by the linear momentum balance, and reads
\begin{equation}\label{eq:chedckeq}
\epsilon_{equilibrium}=\frac{|| \nabla \cdot \left(\boldsymbol{\sigma}\right)||_{L2}}{||\left<\boldsymbol{\sigma}\right>||},
\end{equation}
where $||.||$ denotes the Frobenius norm of a second order tensor and $||.||_{L2}$ the $L_2$ norm of a tensor field. The equilibrium residual tolerance is set to $10^{-8}$ for all the tests presented here. The second residual quantifies the incompatibility of the strain field and is defined as
\begin{equation}\label{eq:chedckcomp}
\epsilon_{compatibility}=\frac{\mathrm{max}\left(\left|\nabla \times\left(\nabla \times\left(\boldsymbol{\varepsilon}\right)\right)\right|\right)}{||\left<\boldsymbol{\varepsilon}\right>||}
\end{equation}
where $\nabla \times$ denotes the rotational operation, $|.|$ is the absolute value of the field and $\mathrm{max}$ extracts the maximum among all the components of all the tensors forming the discrete field. The last error measure is the loading residual, which checks if the boundary conditions are fulfilled, and is defined by
\begin{equation}\label{eq:chedckimp}
\epsilon_{loading}=\frac{||\left<\boldsymbol{\varepsilon}\right>-\overline{\boldsymbol{\varepsilon}}||}{||\overline{\boldsymbol{\varepsilon}}||}
\end{equation}
Under stress control (the macroscopic stress is prescribed), the residual is computed with the stress field $\boldsymbol{\sigma}$ and the prescribed macroscopic stress $\overline{\boldsymbol{\sigma}}$ instead. The compatibility check and the loading check tolerance are established as $10^{-10}$.

Finally, differences in the solution between the two methods are compared by computing the norm of the difference of the solution fields, that follows
\begin{equation}\label{eq:L2}
\text{Diff}(f)=\frac{||f_{DBFFT}-f_{variational}||_{L2}}{||f_{variational}||_{L2}}
\end{equation}
for a generic field $f$ that can be a vector or a tensor field.

\subsection{Linear elastic with strain control: spherical inclusion}
In the first numerical benchmark an idealized linear elastic composite with a regular array of spherical inclusions with variable phase contrast is studied. The RVE corresponds to a cubic periodic domain composed by a spherical inclusion embedded in the matrix,  occupying the $20\%$ volume fraction and discretized in $63\cdot 63 \cdot 63$ voxels. The problem is solved prescribing the full macroscopic strain tensor. The elastic properties of each phase is set to $E_m=70MPa$ and $\nu=0.3$ for the matrix, while for the inclusion the elastic modulus is $E_m=k\cdot E_m$ and $\nu=0.3$, where $k$ is the stiffness contrast. The macroscopic strain applied is uniaxial strain, being $0.1\%$ in loading direction and $0$ for the rest of the components. The linear system given by eq. (\ref{eq:lineqf}) and using the preconditioner defined in eqs. (\ref{eq:pred2}) and (\ref{eq:pred3})  is solved using the Conjugate Gradient method, and at each iterations the three error estimations (eqs. \ref{eq:chedckeq}-\ref{eq:chedckimp}) are computed. The iterative process is performed until reaching a tolerances for equilibrium $\epsilon_{equilibrium}<10^{-8}$ and $10^{-10}$ for the other two estimations or when the number of iterations reached 10$^4$. Ten different cases have been simulated with phase contrasts ranging from very soft inclusions (nearly voids, $k=10^{-5}$) to stiff inclusions ($k=10^{4}$).

\begin{figure}[H]
\includegraphics[width=.49\textwidth]{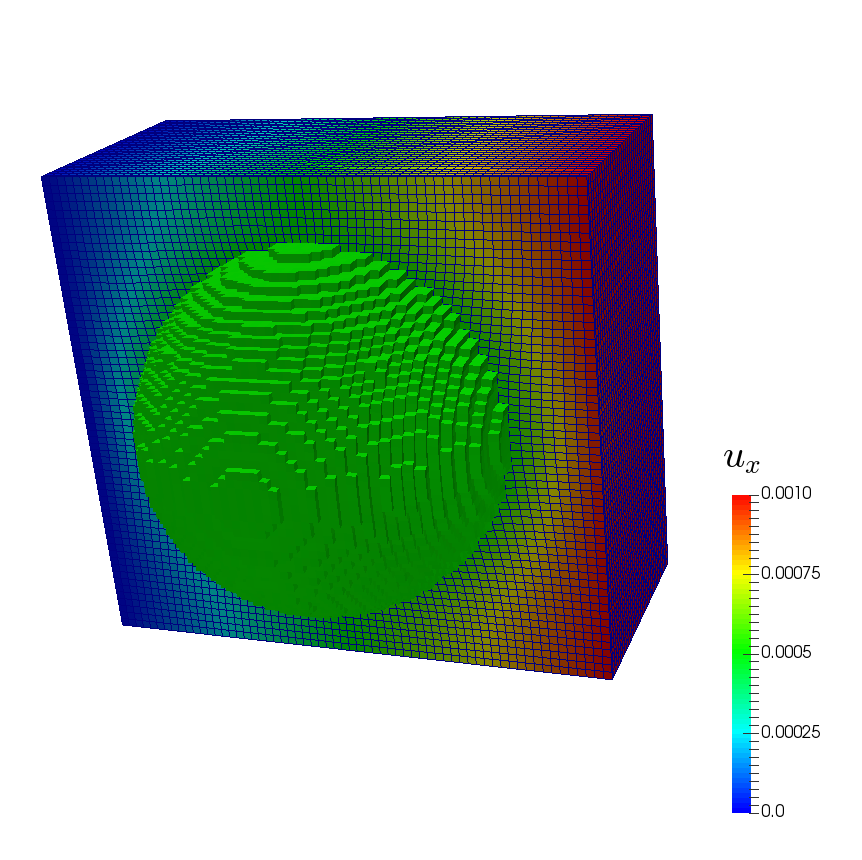}
\includegraphics[width=.49\textwidth]{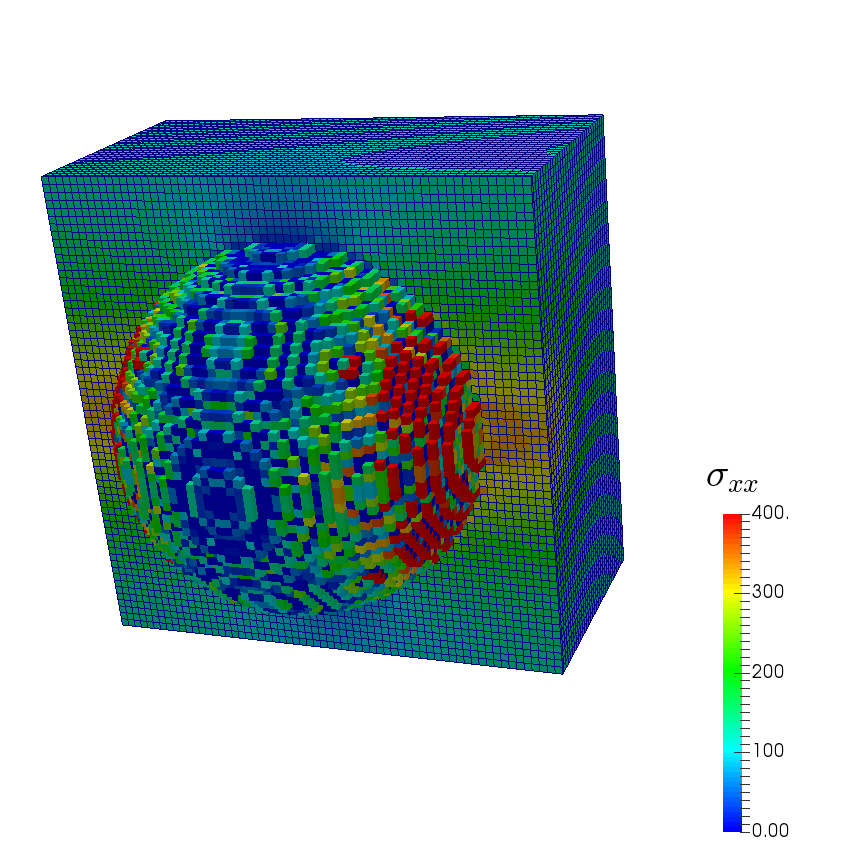}
\caption{Deformed shape ($\times 20$) obtained using the displacement field solved directly in DBFFT (left) and stress field in the loading direction of the elastic inclusion model (right) for phase contrast 10$^5$\%}
\label{fig:geo}
\end{figure}

To illustrate the results obtained and the ability of the DBFFT method to obtain the deformed shape without any further postprocessing, the deformed shape and the axial stress component in the direction of the applied load are shown in Figure \ref{fig:geo} for the case of $k=10^{5}$. Quantitatively, for all the stiffness contrasts, the strain fields obtained with the DBFFT and the variational approach solution are virtually identical, being the difference between these fields computed using eq. \ref{eq:L2} below the equilibrium tolerance ($10^{-8}$). The homogeneized response is therefore also identical, within the numerical precision.

\begin{figure}[H]
\includegraphics[width=.49\textwidth]{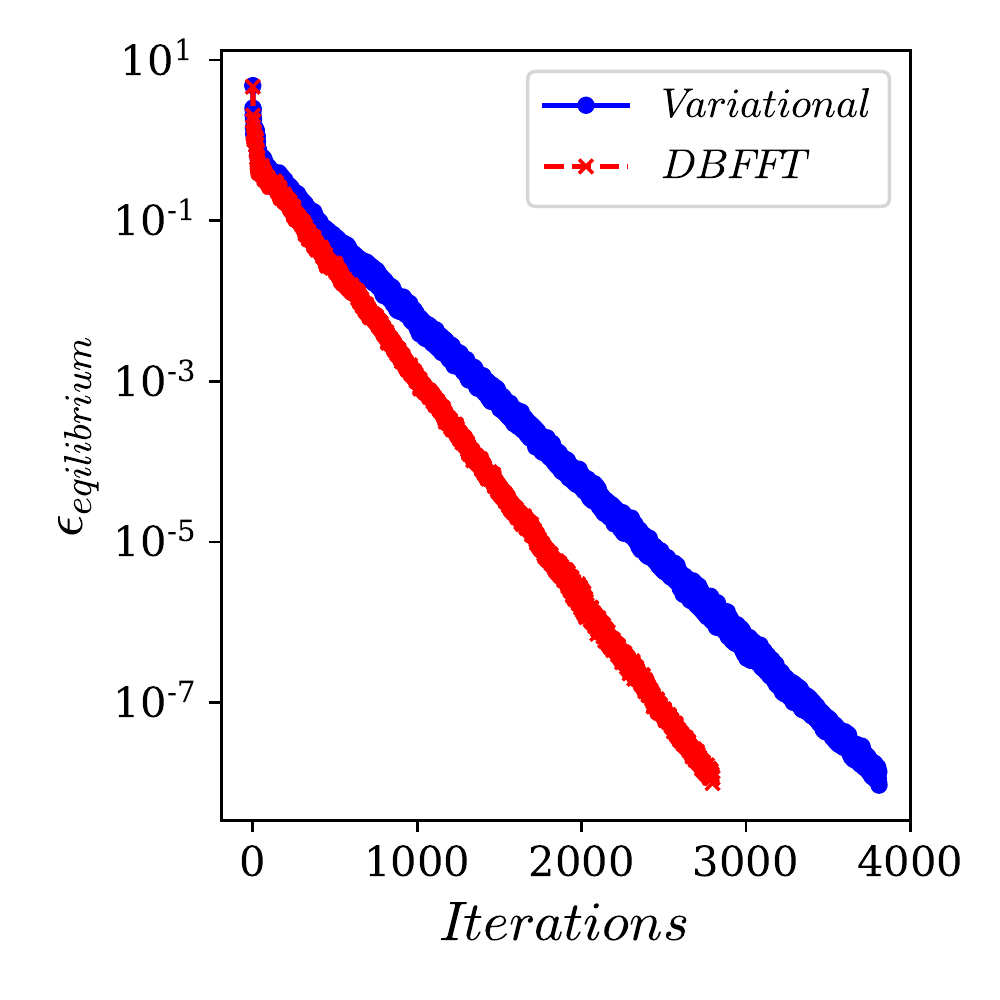}
\includegraphics[width=.49\textwidth]{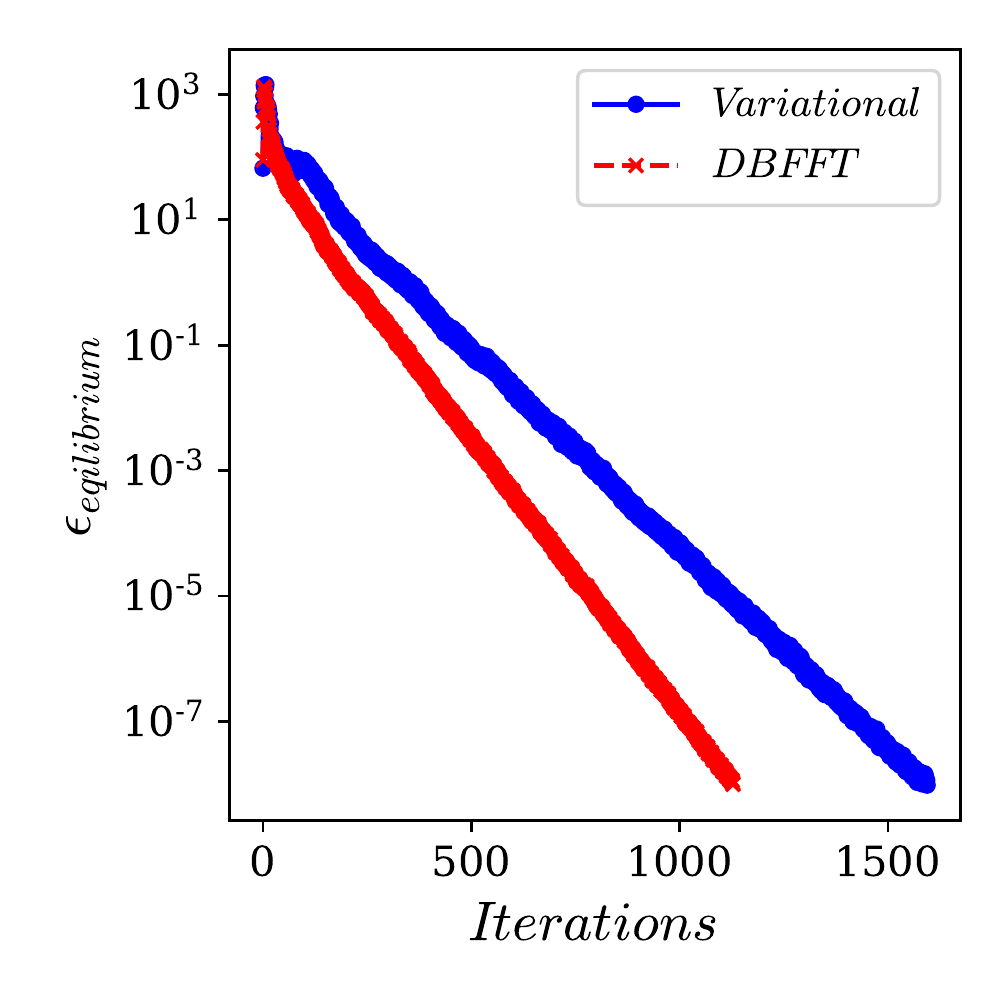}
\caption{Equilibrium residual for the elastic inclusion model when $k=10^{-5}$ (left) and $k=10^{4}$}
\label{fig:div}
\end{figure}

When checking the convergence rates, a remarkable difference is found between the two methods, as shown in Fig. \ref{fig:div} where the residual decay is represented as function of the number of iterations for the cases with the extremal phase contrast, $k=10^{-5}$ and $k=10^{5}$. In all the cases considered, the convergence rate of the DBFFT for the same tolerances and linear solver (conjugate gradient)  is superior, leading to an improvement in efficiency respect to the variational approach that, for the extreme cases $k=10^{-5}$ and $k=10^{5}$, result in a reduction of the number of iterations up to 40\%. It must be noted that the compatibility and loading fulfilment are always accomplished for both methods .

The improvement in the number of iterations is related to the condition number of the matrix representing the linear system of equations. In the case of the DBFFT, the use of a preconditioner reduces the condition number of the matrix and improves the convergence rate allowing to find the solution using less iterations.

\subsection{Linear elastic with stress control: spherical inclusion}
\begin{figure}[H]
\includegraphics[width=.49\textwidth]{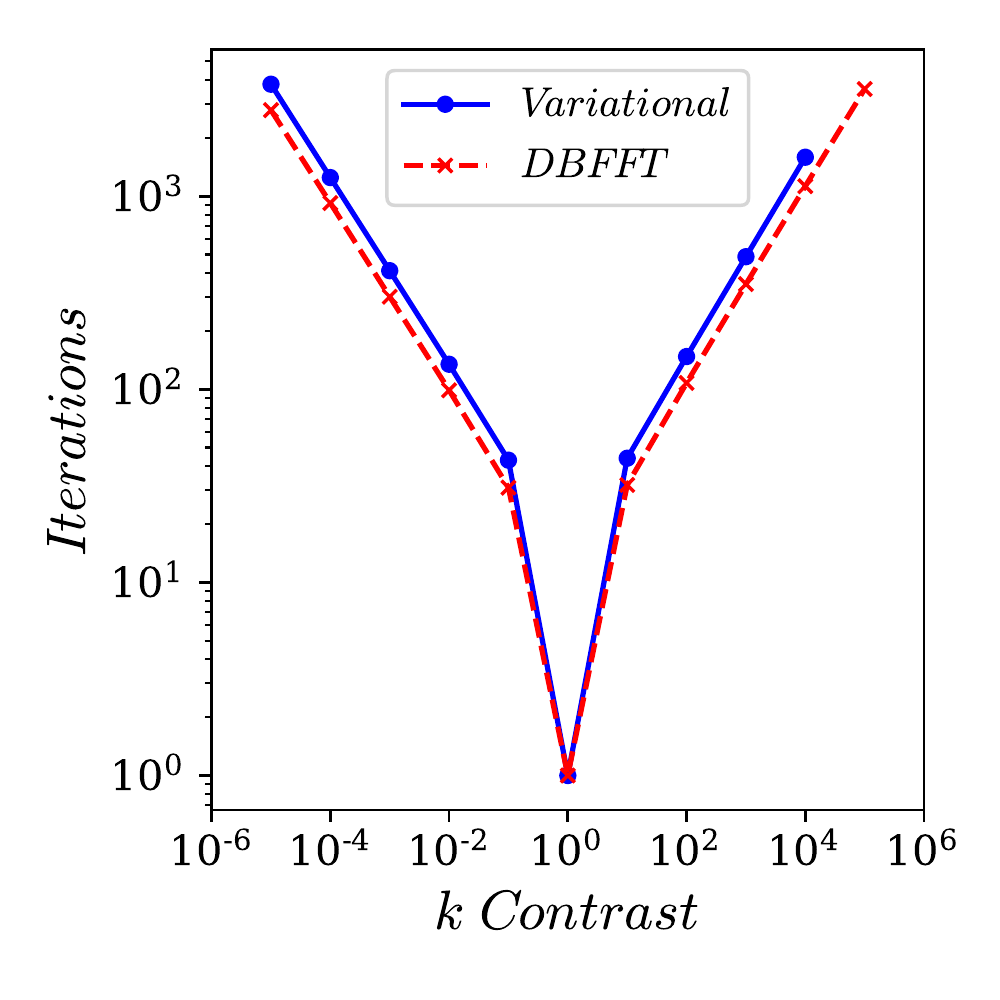}
\includegraphics[width=.49\textwidth]{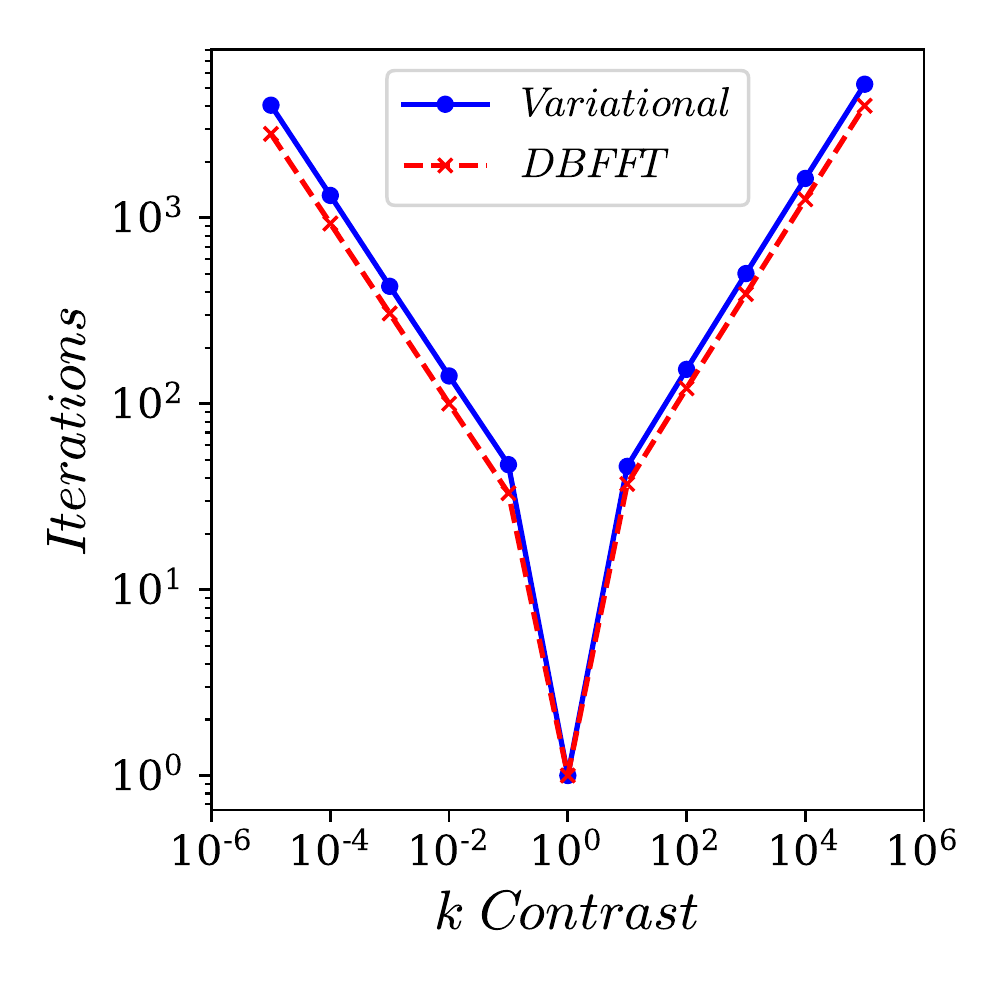}
\caption{Number of iterations as a function of the stiffness contrast for elastic linear inclusion: strain control (left) and stress control (right)}
\label{fig:contrast}
\end{figure}

In this section the results of the DBFFT method using macroscopic stress control are assessed solving the same case of the elastic inclusion presented in the previous section. RVE shape, material properties and tolerances are the same but in this case the load applied is a uniaxial macroscopic stress in the loading direction, being the rest of the components $0$. As in the previous case, the differences in stress and strain fields between the DBFFT with stress control and the stress control version of the variational scheme proposed in \cite{Lucarini2019b} are below the equilibrium tolerance. Respect the efficiency, a figure summarizing the results obtained for all the stiffness contrasts considered is represented in Figure \ref{fig:contrast} (left) where the number of iterations for both methods is represented as function of the phase contrast. In Fig. \ref{fig:contrast} (right) the results obtained in the previous section are represented in the same type of diagram for a better comparison. As it happens under strain control, here in all the cases the compatibility and loading error estimators are fulfilled for both methods. The convergence rates are also higher for the DBFFT algorithm and depend on the stiffness contrast as shown in Figure \ref{fig:contrast}. The comparison in number of iterations needed to reach the equilibrium tolerance is almost identical to the strain controlled simulations, a reduction near 40\% in the case of compliant inclusions.

\subsection{Non linear simulation with hyperelastic materials: random array of spherical voids}
\begin{figure}[H]
\includegraphics[width=.49\textwidth]{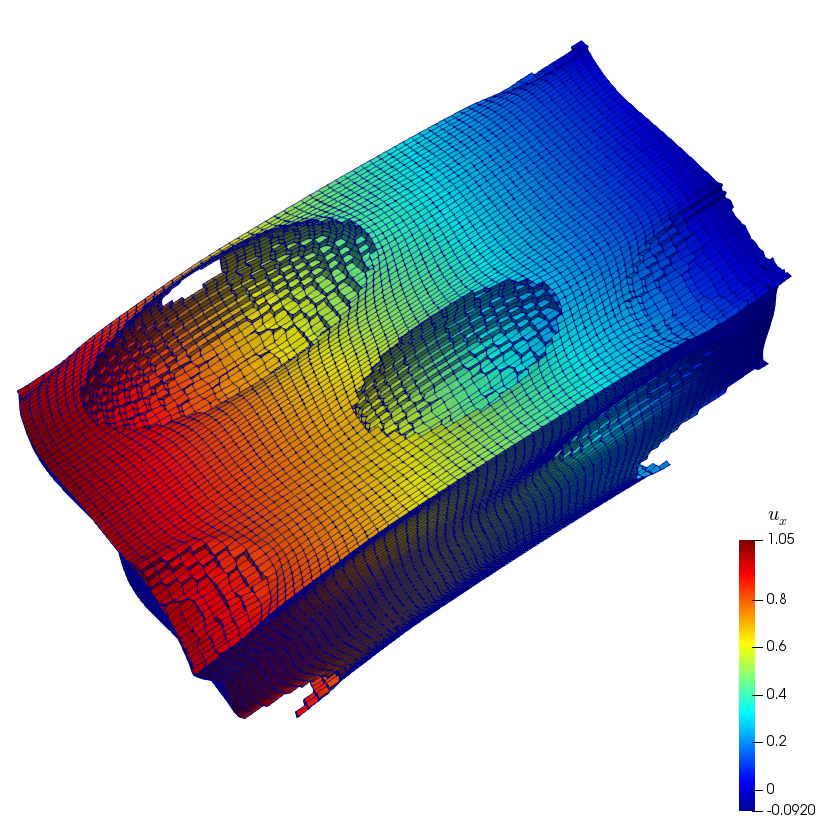}
\includegraphics[width=.49\textwidth]{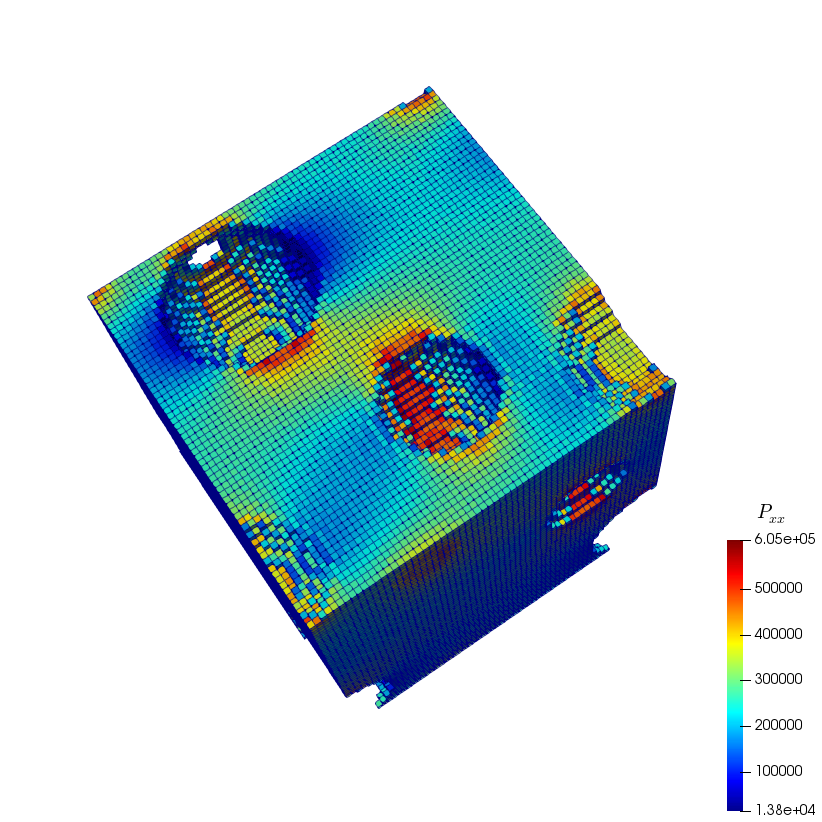}
\caption{Deformed shape with the displacement field (left) and first Piola-Kichhoff stress field in the loading direction (right) of the porous matrix}
\label{fig:hypgeo}
\end{figure}

The third benchmark consists in a non-linear hyperelastic matrix containing a random distribution of 5 spherical pores and a total porosity volume fraction of 0.2. Opposite to FE, in FFT approaches voids cannot be just removed from the grid since Fourier transforms and differential operators act on full fields defined in regular grid in a hexaedrical domain. For this reason, voids are usually considered as very compliant elastic inclusions, the same approach followed here. The RVE considered is a cube containing $63^3$ voxels (Fig. \ref{fig:hypgeo}).The test simulated consists in applying a uniaxial elongation $\lambda=\frac{L}{L_0}=2$, being $L$ and $L_0$ the final and initial lengths of the cell in the loading direction, and keeping the other directions undeformed. Both the matrix and the compliant inclusions follow a Saint Venant-Kirchhoff hyperelastic model. The matrix properties are $E=70MPa$ and $\nu =0.3$ (Young and Poisson modulus) and the stiffness of the inclusions is 100 times smaller. Load is homogeneously divided in 5 load increments and the equilibrium is solved in each of them using Newton method, as described in the model description. The tolerance of the Newton iterations is 10$^{-6}$ and the three residuals proposed for the linear simulations are adapted to the non-linear case and used to check the convergence of the linear iterations. In particular, the equilibrium error estimator is computed modifying eq. \ref{eq:chedckeq} by replacing the stress with the first Piola-Kirchhoff stress and the divergence with the divergence in the reference configuration. The compatibility check is also adapted to finite strain, being computed as

\begin{equation}
\epsilon_{compatibility}=\frac{\mathrm{max}\left(|\nabla_0 \times(\mathbf{F})|\right)}{||\left<\mathbf{F}\right> -\mathbf{I}||}
\end{equation}
where $\nabla_0 \times$ represents the rotational of a tensor field in the reference configuration.

The deformed shape obtained using directly the displacement obtained by the DBFFT method is represented in Fig. \ref{fig:hypgeo} without any magnification. In the same figure, on the right, the normal component of the first Piola-Kichhoff stress field in the loading direction is represented for the porous matrix. In the Figure \ref{fig:hypgeo} it can be observed how the pore shape evolves from spherical ellipsoidal and how the stress in the loading direction is concentrated in the void diameter in planes normal to the loading direction. Respect the accuracy of the solution, the provided by DBFFT is again almost identical to the one obtained by the variational method, being the differences (eq. \ref{eq:L2}) in both deformation gradient and stress fields below the equilibrium tolerance ($10^{-8}$) at the end of each increment. The compatibility and loading requirements are always fulfilled.

\begin{figure}[H]
\includegraphics[width=.49\textwidth]{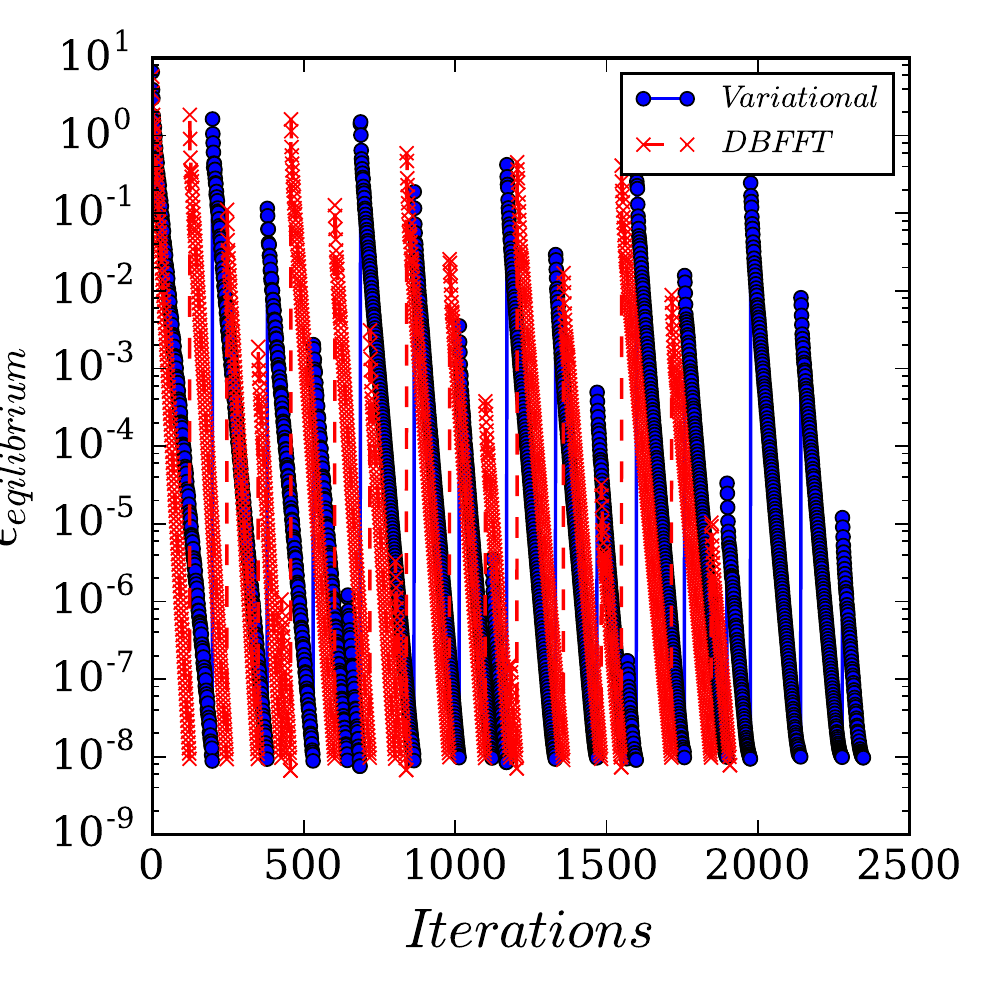}
\includegraphics[width=.49\textwidth]{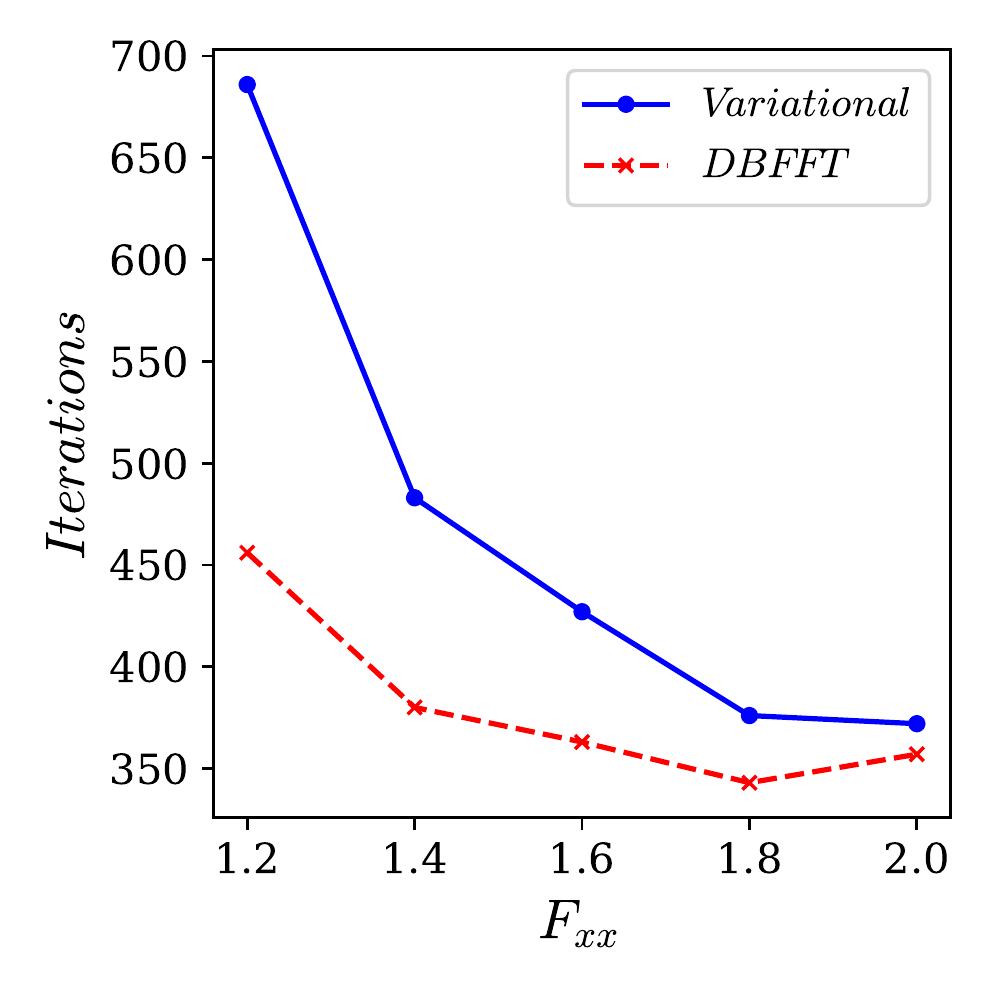}
\caption{Equilibrium residual (left) and number of iterations per increment (right) for the hyperelastic porous matrix model}
\label{fig:hypdiv}
\end{figure}

Finally, the efficiency of the method for this non-linear elastic case is analyzed representing the evolution of the equilibrium error estimator (eq. \ref{eq:lineqf}) as function of the total iteration number (Fig. \ref{fig:hypdiv}, left), and the total number of iterations per increment at the end of each strain increment (Fig. \ref{fig:hypdiv}, right). The number of iteration in both curves refers to the total number of iterations performed by the conjugate gradient gradient considering all the Newton iterations. From Fig. \ref{fig:hypdiv}(left) it can be observed that DBFFT converges faster and requires less linear iterations per Newton iteration to reach the tolerance resulting in a smaller number of increments for the simulation than the variational method. The total iteration number is reduced by $20\%$, which is translated in a computational time saving of similar order. It must be noted that the better performance of the DBFFT respect to the variational scheme is reduced in the last increment, at a very large strain. The reason for this reduction in the convergence rate is that the preconditioner used, computed with the initial tangent stiffness, does not improve the condition number of the system for very distorted RVEs.

\subsection{Non linear with path-dependency: elasto-visco-plastic polycrystal}
\begin{figure}[H]
\includegraphics[width=.49\textwidth]{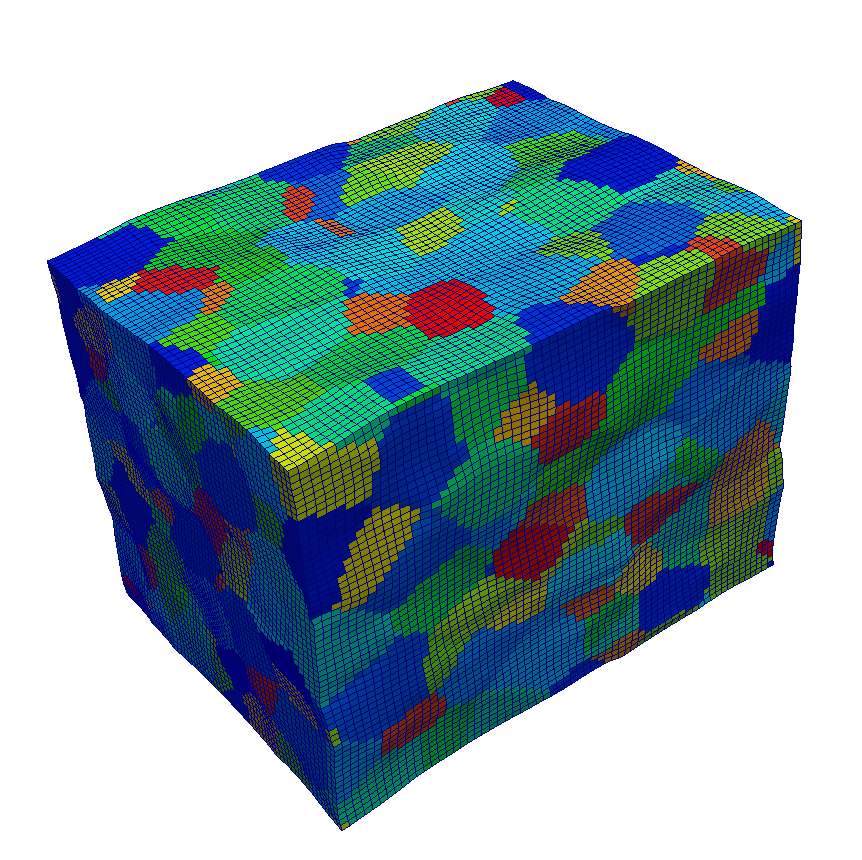}
\includegraphics[width=.49\textwidth]{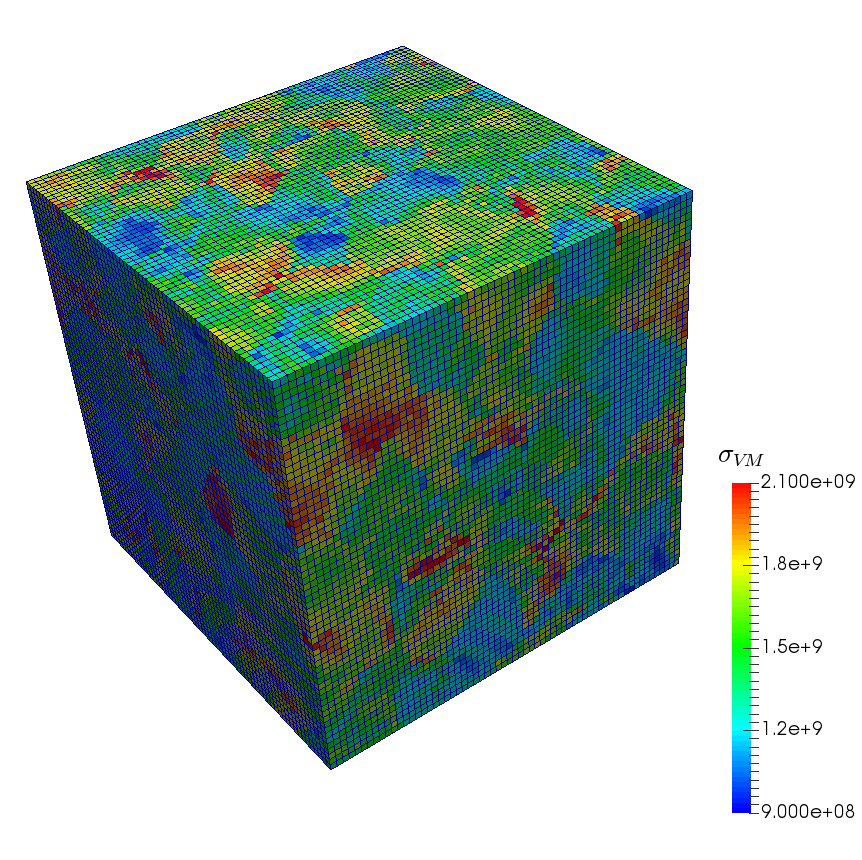}
\caption{Deformed shape ($\times$10) showing grains (left) and Von Mises stress field (right) of the polycrystal}
\label{fig:polydef}
\end{figure}

The last benchmark proposed is a metallic polycrystal, a material class in which FFT homogenization has been widely employed \cite{L01,SEGURADO2018}. This is an ideal case because it illustrates the behavior of a material with loading path and strain rate dependencies. The RVE is  a cube discretized in $63^3$ voxels and the model contains around 200 randomly oriented grains (Fig. \ref{fig:polydef}). Grain distribution in the RVE is generated using a Voronoi tessellation and grain orientations are random. Each grain follows a crystal plasticity (CP) model with elasto-visco-plastic behavior. The CP model represents  a Ni-based FCC superalloy with 12 octaedrical slip systems. The model and the parameters used are taken from \citep{CRUZADO2015242}. Strain control is used prescribing the full deformation gradient. A final elongation of $\lambda=0.02$ is imposed in $400$s in the loading direction and a negative strain is applied in transverse direction $\lambda'=1/\sqrt{\lambda}$ to produce a final isochoric deformation. Shear components of the deformation gradient are set to 0 and time is discretized in $20$ regular increments of $20s$.

The DBFFT method is able to finish the simulation using the standard error tests, the equilibrium estimator given by (eq. \ref{eq:lineqf}) using the Piola stress and $10^{-8}$ as value of the tolerance. On the contrary, the variational approach after a few increments cease to converge and the error estimator is stacked near the tolerance limit. This lack of convergence for very small tolerances in the case of the polycrystal is due to the ill-condition of the linear operator, that might be caused by the large differences in the local material tangents during the elastic-plastic transition.  For this reason, the residual proposed in \citep{DEGEUS2017} is used instead for the equilibrium check and, in this case, both methods are able to converge. The norm of the difference of deformation gradient field is $5\cdot 10^{-7}$ and the loading and compatibility check are fulfilled in both methods.

The deformed shape obtained with the displacements obtained with DBFFT method is represented in the left part of Figure \ref{fig:polydef} using a magnification of 10. A smooth deformation of the surface is observed corresponding to the different effective stiffness of the grains due to their orientation. The Von Mises stress invariant of the Cauchy stress is shown in the right image of Figure \ref{fig:polydef}. In this case, the concentration of the stress in certain grains can be clearly observed. This stress concentration is due to the orientation of the grain as well as the orientation of the surrounding grains.

The convergence is analyzed representing in Figure \ref{fig:polydef} the evolution of the equilibrium error estimator (defined in \cite{DEGEUS2017}) as function of the total iteration number (left) and the total number of iterations per increment at the end of each strain increment (right). The DBFFT performance is in this case clearly superior to the variational approach. The total number of iterations for the DBFTT is reduced by $60\%$ even considering a tighter equilibrium residual tolerance.
Respect to the evolution of the number of iterations per increment, it shows different stages for both methods. In the first increments, corresponding to the elastic response, the number of iterations is very small due to the limited phase contrast. Then, number increases very fast and is kept high (around 200 in the variational and around 70 in the DBFFT) in a strain range corresponding to the elasto-plastic transition. This is due to the very high stiffness contrast at this stage caused by the presence of elastic points with very large stiffness with points fully plastified with tangent stiffness orders of magnitude smaller. After this transition, the number of iterations decreases again and this stage corresponds to te fully plastic regime. Note that during the fully plastic regime the number of iterations per increment in the DBFFT method is kept constant while grows linearly in the variational method.

\begin{figure}[H]
\includegraphics[width=.49\textwidth]{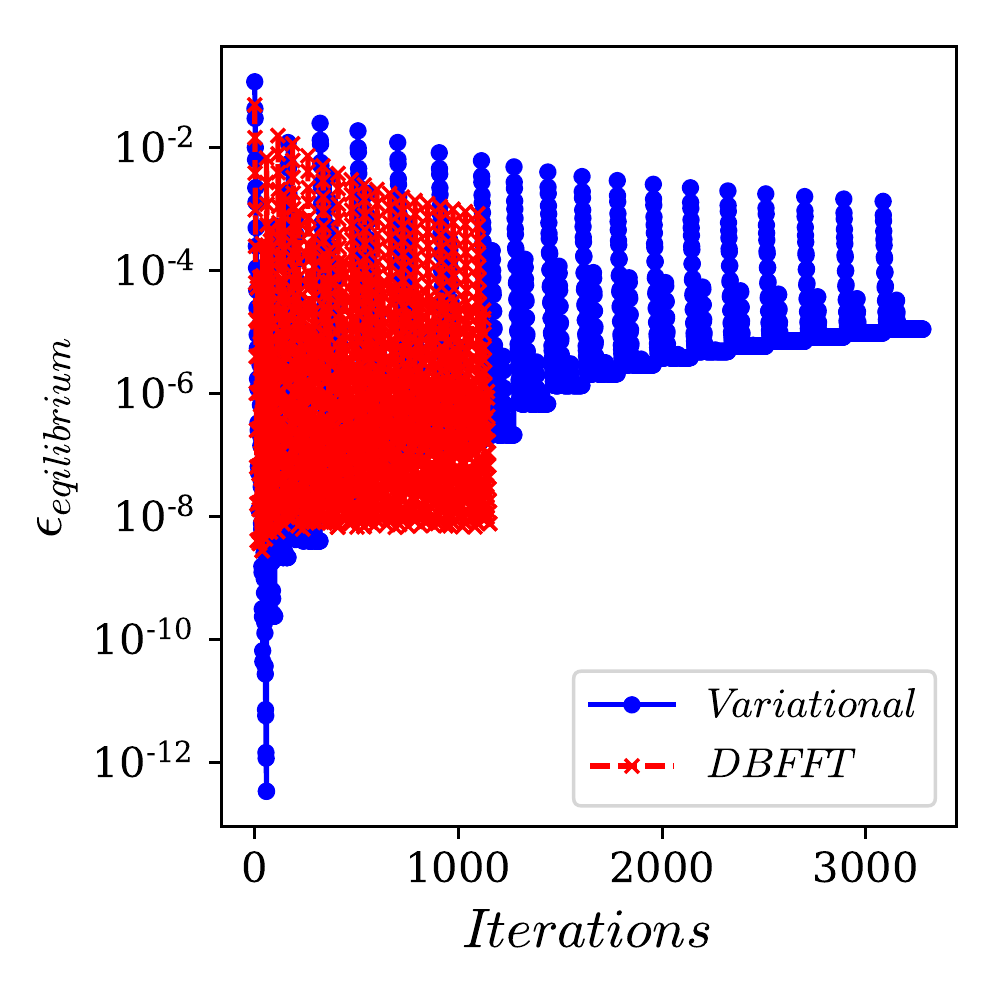}
\includegraphics[width=.49\textwidth]{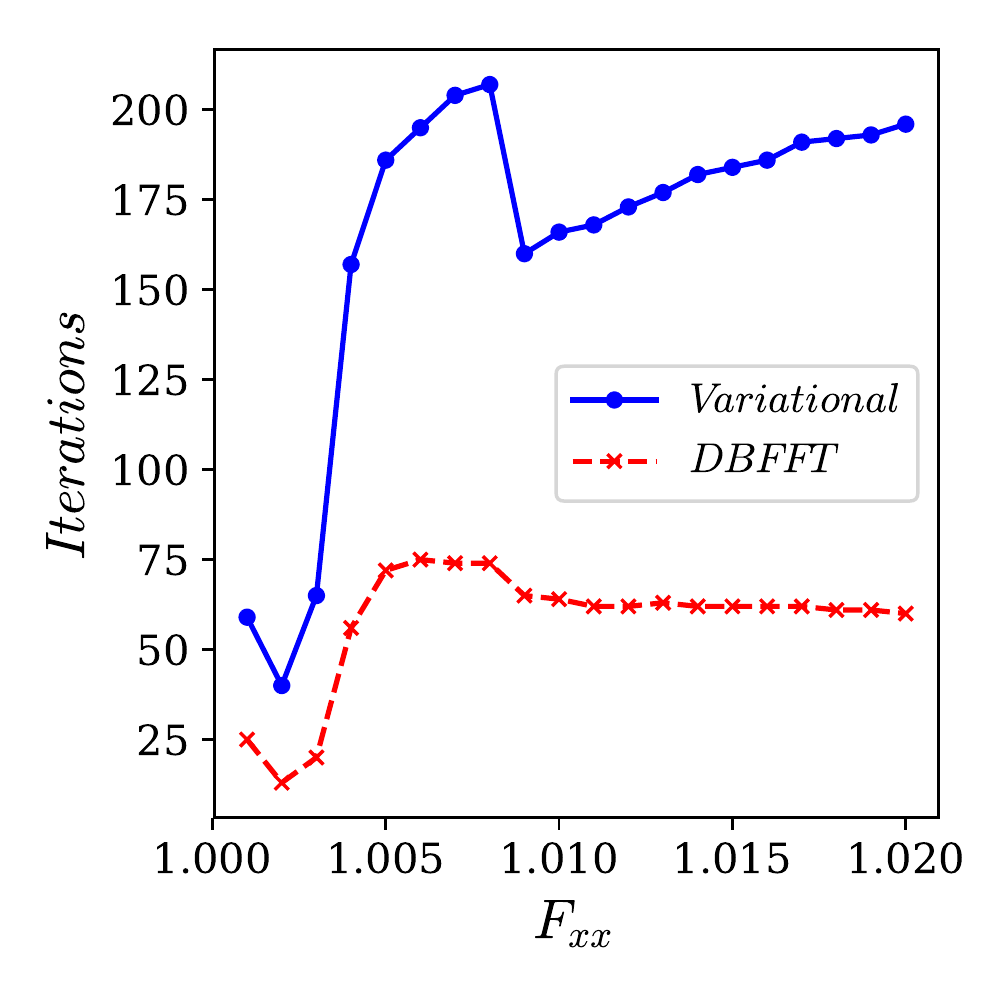}
\caption{Equilibrium residual (left) and number of iterations per increment (right) for the polysrystal}
\label{fig:poldiv}
\end{figure}

\section{Discussion}
The results obtained shows in all the cases examined an improvement in the convergence rate of the DBFFT method compared to the variational scheme. This leads to a reduction in the computational time proportional to the number of iterations reduced since the heaviest operations for each equilibrium iteration, the direct and inverse Fourier transform of a second order tensor field, are the same in both methods. The rest of operations performed in the conjugate gradient do not represent an important time contribution to the total but DBFFT also presents a small computational saving in this case. This improvement in the conjugate gradient are due to the reduction of the number of components of the solution vector in the system,  $1/2$ (small strains) or $1/3$ (finite strains) the size of the methods in which the unknown variable is the strain.

The dependency of the convergence with the stiffness contrast is related to the condition number of the linear system of equations that increases with the phase contrast. For this reason, preconditioners are fundamental to improve the convergence. This is the main reason why the DBFFT needs less iteration to reach the equilibrium than the variational method that does not admit preconditioners. Moreover, this improvement has been obtained with the preconditioner proposed in this work, but other preconditioners could be found that further improve this behavior.

Regarding memory allocation needs, the variational approach need to store $5$ tensor fields while in DBFFT $6$ vector fields and $1$ tensor field are saved. It can be remarked that in the new method, two third order tensor fields defined in frequencies are used while in the rest of FFT schemes a forth order tensor needs to be computed, and saved to avoid recomputating them. These considerations imply a reduction of memory around 30\%.

\section{Conclusions}
An efficient and robust FFT framework for  computational homogenization, the DBFFT method, is proposed. The method uses the displacement/deformation gradient field as unknown variable and the preconditioned conjugate gradient as linear equation solver in the Fourier space. The framework allows the simulation of materials with both linear and non-linear constitutive equations under both small strains and finite deformations. The DBFFT includes also a method for direct an efficient stress and mixed control.

It is shown first that the algorithm reproduces the same results (within numerical precision) than other FFT schemes.  The main advantages of the method are
\begin{itemize}
   \item The algorithm does not require the definition of a reference medium
   \item The linear systems in which the method arises are fully ranked and therefore admits the use of the conjugate gradient as method, which shows good convergence properties compared to fixed point iteration methods, and preconditioners which improve the convergence rate
   \item In the iterative solution procedure, the convergence rate is higher than the FFT based variational approach, being in some cases up to 40\% the reduction
   \item The memory allocation necessity is about 30\% lower than in the rest of the FFT-based schemes
   \item The number of operations needed for each iteration is slightly lower
   \item The displacement field is computed directly and should not be reconstructed in the postprocessing
\end{itemize}

It should be noted finally that this performance improvements are based on the preconditioner proposed but the efficiency of the scheme could be further improved in future work by the developments of new ad-hoc preconditioners.

\section*{Acknowledgments}
This investigation was supported by ITP Aero, the Spanish Ministry of Economy and Competitiveness through the project DPI2015-67667-C3-2-R and the Spanish Science, Innovation and Universities through the project NVIDIA (RTC-2017-6150-4).

\bibliographystyle{unsrt}
%\bibliography{bibliography}
\end{document}